\documentclass[
10pt,aps,
notitlepage, 
bibnotes,
pra,
twocolumn,
showpacs,
preprintnumbers,
amsmath,
amssymb,
floatfix
]{revtex4-2}

\usepackage{graphicx}
\usepackage{dsfont}
\usepackage{bm}
\usepackage{color}
\usepackage{xcolor}
\usepackage{braket}
\usepackage[bookmarksnumbered,pdfpagelabels=true,plainpages=false,colorlinks=true,linkcolor=blue,citecolor=red,urlcolor=blue]{hyperref}
\usepackage[capitalise]{cleveref}

\begin{document}

\def\afflux{Department of Physics and Materials Science, University of Luxembourg, L-1511 Luxembourg, Luxembourg}

\title{Skyrmionic Schr\"odinger cat states in monoaxial chiral magnets}

\author{\v{S}tefan Li\v{s}\v{c}\'{a}k}
\email{stefan.liscak@uni.lu}
\thanks{These authors contributed equally.}
\affiliation{\afflux}

\author{Andreas Haller}
\thanks{These authors contributed equally.}
\affiliation{\afflux}

\author{Andreas Michels}
\affiliation{\afflux}

\author{Thomas L. Schmidt}
\affiliation{\afflux}

\author{Vladyslav M. Kuchkin}
\affiliation{\afflux}

\begin{abstract}
    We study the low-energy excitation spectra of a spin-1/2 quantum Heisenberg model with a monoaxial Dzyaloshinskii-Moriya interaction.
    Using the density matrix renormalization group method, our analysis reveals a degeneracy between skyrmion and antiskyrmion states, enabling the formation of a mesoscopic Schrödinger cat state---a quantum superposition of these topologically distinct textures.
    To characterize this nontrivial state, we compute two-point spin correlation functions, highlighting signatures accessible via neutron scattering experiments.
    Furthermore, we demonstrate that applying a magnetic field gradient induces a coherent time evolution of the cat state, offering a controllable mechanism for its manipulation.
    These findings provide a framework for the detection of skyrmionic Schrödinger cat states in quantum magnets.
\end{abstract}

\date{\today}

\maketitle

%%%%%%%%%%
\section{Introduction}
%%%%%%%%%%

Two-dimensional skyrmions are topological solitons with particle-like properties, holding significant potential for applications in spintronics~\cite{Tokura2020}, magnetic memory recording, neuromorphic computing~\cite{Song2020}, and quantum computing~\cite{Psaroudaki2023,Xia2023,Petrovic2024,Chudnovsky2025}.
Initially, skyrmions were studied within the framework of a classical chiral magnet Hamiltonian~\cite{Bogdanov_89, Bogdanov_1994}, which consists of competing Heisenberg exchange and Dzyaloshinskii-Moriya interactions~\cite{Dzyaloshinskii, Moriya} (DMI).
Recently, however, the quantum properties of skyrmions have garnered significant attention~\cite{Ochoa2019, Lohani2019, Sotnikov2021, Diaz2021, Haller2022, Siegl2022, Vijayan2023, Mazurenko2023, Peters2023, Joshi2023, Sotnikov2023, Joshi2024, Salvati2024, Sorn2024, Zhao2024, Makuta2024}, particularly within the context of quantum spin Hamiltonians, due to their potential use as qubits in quantum computation~\cite{Psaroudaki2023, Xia2023,Chudnovsky2025}, as carriers of topological bound modes~\cite{Diaz2021_2,Nothhelfer2022}, or as catalysts for topological superconductivity~\cite{Maeland2022_1,Maeland2022_2}.

\begin{figure*}[ht]
    \centering
    \includegraphics[width=17.7cm]{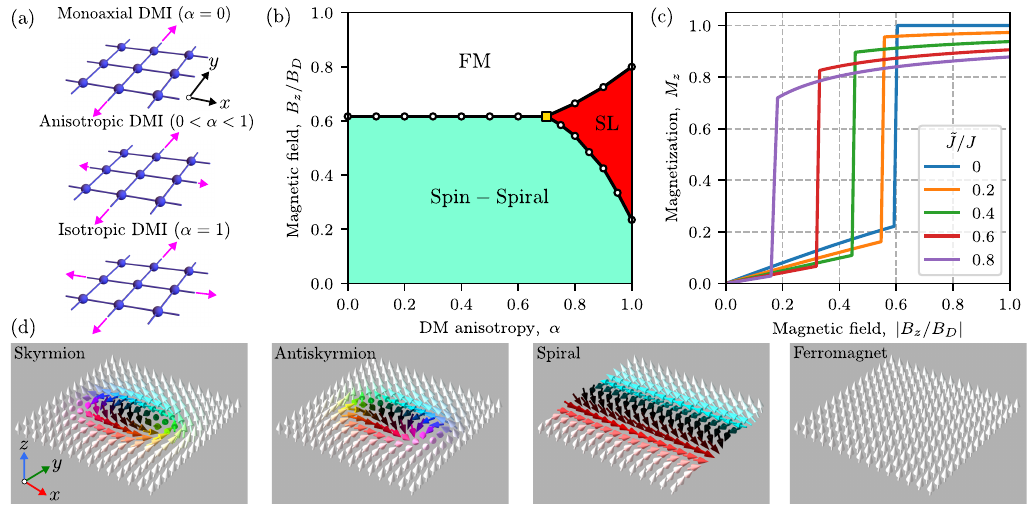}
    \caption{
    Panel~(a) shows different types of Bloch-like DMI: isotropic, anisotropic, and monoaxial, which can be distinguished by the parameter $\alpha$ in Eq.~(2) in the main text.
    Panel~(b) shows the phase diagram in terms of $\alpha$ and the external field $B_{z}/B_\mathrm{D}$, and contains three phases: saturated ferromagnetic (FM), spin-spiral and skyrmion lattice (SL) states.
    Panel~(c) shows the transitions between spin-spiral and FM states for different values of magnetic field and easy-plane anisotropies $\tilde{J}$. 
    Panel~(d) displays the magnetisation textures relevant to the phase landscape in the monoaxial magnet, namely the (anti)skyrmion, spin-spiral and ferromagnet.
    The arrow orientations are color-coded based on the azimuthal and polar angles.
    White (black) arrows associate with spins pointing up (down) with respect to the plane.
    } 
    \label{fig:mono_texture}
\end{figure*}

Unlike their classical counterparts, quantum skyrmions are quasiparticles that exhibit novel physical phenomena arising from the entanglement properties of the underlying spin system, usually spin-1/2 particles.
In this work, we investigate quantum skyrmions in materials characterized by highly anisotropic DMI along a single spatial dimension while the Heisenberg exchange interaction remains isotropic.
These materials, known as monoaxial chiral magnets~\cite{Togawa2012}, are of particular interest due to their unique ability to stabilize both skyrmion and antiskyrmion states with equal energy~\cite{Kuchkin2023}, in contrast to isotropic chiral magnets, where the antiskyrmion~\cite{Kuchkin2020, Zheng2022} always represents a higher energy state.
From a quantum perspective, such skyrmions and antiskyrmions can thus form a mesoscopic superposition state, dubbed Schr\"odinger cat state, which cannot be described by classical micromagnetic models.
It is worth noting that the term ``Schr\"odinger's cat'' concerning a magnetic skyrmion in narrow racetrack devices has been recently introduced by Chudnovsky and Garanin in Ref.~\cite{Chudnovsky2025}.
We want to stress that, in the present work, we utilize this name for the skyrmion-antiskyrmion superposition.
The state of the system can be described by a wave function $\ket{\psi}$, which we use to calculate the magnetization vector field as an expectation value $\bm m=\braket{\psi|\hat{\bm S}|\psi}/\hbar$, of the spin operators $\hat{\bm S}$.
In a classical case, the magnetization has a fixed length $|\bm m|=\mathrm{const}$, while in the quantum case, it can deviate and even completely vanish.

In this work, we present our findings on the quantum properties of monoaxial skyrmions and antiskyrmions, and we provide a pathway towards their experimental observation through polarized neutron scattering.
Furthermore, by solving the corresponding Schr\"odinger equation, we investigate the dynamics of a skyrmion-antiskyrmion superposition state in an external magnetic gradient field.
By analyzing these dynamics in detail, we compare our results with predictions from simulations based on the classical Landau-Lifshitz and Thiele equations, thereby establishing a connection between the quantum Heisenberg model and micromagnetic theory in monoaxial chiral magnet systems.

The paper is structured as follows: In \cref{sec:model}, we introduce the quantum spin model used to describe monoaxial chiral magnets and we discuss its classical counterpart, including the phase diagram.
\Cref{sec:results} contains the main results of our paper, covering both static and dynamic properties of quantum skyrmions, antiskyrmions, and their superpositions, organized into several focused subsections.
Section~\ref{summary} concludes the paper by summarizing the main findings of this study.
Technical details and extended derivations are provided in Appendices~\ref{sec:product_state_approx}$-$\ref{sec:tdvp}.

%%%%%%%%%%
\section{Model and Methods}
\label{sec:model}
%%%%%%%%%%

We consider the following Heisenberg Hamiltonian
\begin{align}
    \!\!\!s\hat{H}
    =
    -\dfrac{1}{2}\!\sum_{\braket{ij}}\!
    \left[J\hat{\bm S}_i \cdot \hat{\bm S}_j 
    \!+\!
    {\bm D}_{ij} \!\cdot \!\left(\!\hat{\bm S}_i \!\times\! \hat{\bm S}_j\!\right)\!\right]
    \!-\!
    \sum_i \bm B_i\cdot\hat{\bm S}_i
    ,
    \label{eq:hamiltonian}
\end{align}
where $\hat{\bm S}_i$ denotes the vector of spin operators for the spin at position $\bm r_i$, and $\bm B_i =  (B + 2B_{\rm grad}y_i/(N_y-1)) \bm e_z$ is the sum of a uniform and a gradient Zeeman field, and $N_y$ the number of sites along $y$.
First, we investigate the static properties for $B_{\rm grad}=0$.
From now on, we consider the special case $s=1/2$, and $J>0$ (ferromagnetic exchange coupling).
In the monoaxial limit, the DMI vector $\bm D_{ij}=Dr_{ij}^y\bm{e}_{y}$ ($D>0$ is the DMI amplitude) is given by the projection of $\bm{r}_{ij}=(r_{ij}^x,r_{ij}^y)$, the unit displacement vector between the sites $i$ and $j$, onto the $y$ axis.
For some calculations, we will consider a more general anisotropic DMI with anisotropy parameter $\alpha$, so in the most general case we will use $\bm D_{ij}= D\left(\alpha r_{ij}^x\bm{e}_{x}+r_{ij}^y\bm{e}_{y}\right)$ [see~\cref{fig:mono_texture}(a)]. 
The parameter $\alpha\in[-1,1]$ allows us to describe DMI of various types: Bloch or bulk type favoring skyrmions $(\alpha>0)$ and D$_{2\rm d}$ type favoring antiskyrmions $(\alpha<0)$~\cite{Hoffmann2017}.

\Cref{fig:mono_texture}(b) contains the phase diagram in the atomistic spin model -- the classical analog of~\cref{eq:hamiltonian} -- for various DMI anisotropies in the micromagnetic limit.
For these simulations, we optimized various magnetic phases with the conjugate gradient method for several external magnetic fields in units of $B_\mathrm{D}=D^2/2J$ and dimensionless DMI anisotropy $\alpha$.
An interesting feature of the isotropic chiral magnets is their ability to host skyrmion lattices at a certain range of externally applied fields.
In contrast, for monoaxial magnets, only two phases are possible: the ferromagnetic (FM) state and the spin-spiral state.
For intermediate values of the DMI anisotropy ($\alpha\simeq 0.7$) there is a triple point in which the energies of the FM, the spin-spiral, and the skyrmion lattice phases coincide.
In the range $0 \leq \alpha \lesssim 0.7$, magnetic skyrmions represent excited states.

Strictly speaking, monoaxial chiral magnets are characterized by the presence of effective easy-plane anisotropy.
To model such anisotropy in the quantum system described by \cref{eq:hamiltonian}, we consider an additional term $\tilde{J}(\hat S_{i,x} \hat S_{j,x} + \hat S_{i,z} \hat S_{j,z})$ with coupling constant $\tilde{J}<0$.
\cref{fig:mono_texture}(c) shows the influence of magnetocrystalline anisotropy $\tilde{J}$ on the transition between FM and spin-spiral states in the monoaxial case, computed for the ground states of~\cref{eq:hamiltonian}. 
From a symmetry point of view, to study the ground state of a monoaxial chiral magnet, it is enough to consider a one-dimensional chain along the $y$-axis.
Numerically, the Hamiltonian can be solved by exact diagonalization for small systems.
We assume periodic boundary conditions to eliminate boundary effects like the chiral surface twist.
At low magnetic fields, the ground state is a spin spiral, while at stronger fields, the ferromagnetic state is the lowest-energy state.
The transition field depends on the anisotropy $\tilde{J}$ and gradually decreases for higher values of $\tilde{J}$. 
Thus, a nonzero anisotropy $\tilde{J}\neq0$ only influences the value of the critical magnetic field, and in the following we omit it for simplicity and set $\tilde{J}=0$.
Likewise, we neglect long-range dipolar field interactions in Eq.~\eqref{eq:hamiltonian}, as their inclusion leads only to quantitative changes in the results presented below.
In the classical case, dipolar fields break the degeneracy between skyrmion and antiskyrmion states, but this effect can always be compensated for by a proper choice of DMI anisotropy $\alpha$.
As a consequence, the results presented here for monoaxial chiral magnets will also hold for anisotropic DMI materials belonging to the D$_{2\rm d}$ symmetry group.
\begin{figure*}[t!]
    \centering
    \includegraphics[width=\linewidth]{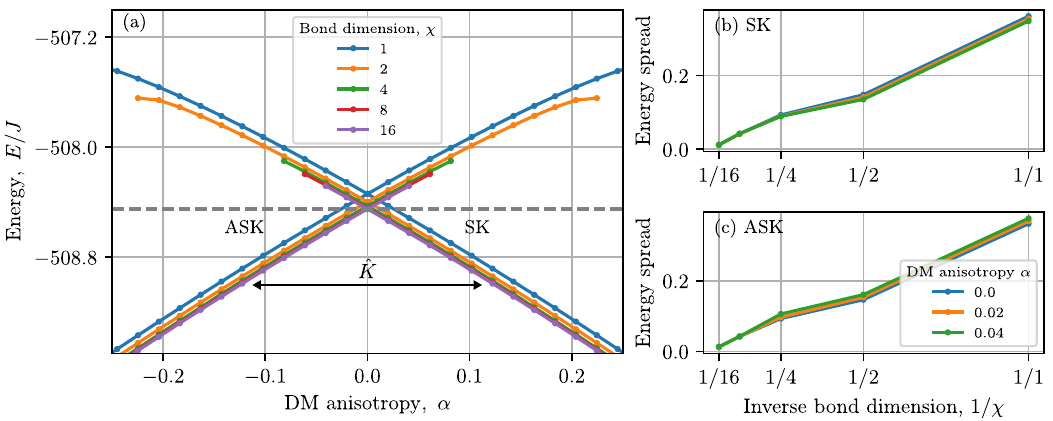}
    \caption{
        Panel~(a) shows energies of skyrmion (SK) and antiskyrmion (ASK) eigenstates as functions of the parameter $\alpha$ at $B_{\rm grad}=0$ for different bond dimensions $\chi$. 
        Note that lines of excited states are getting shorter with increasing alpha, since the antiskyrmion (skyrmion) constitutes an eigenstate of $\hat H$ only in a region $\alpha \leq \alpha_c$ ($\alpha \geq -\alpha_c$). 
        Using a linear extrapolation of the critical $\alpha_c(1/\chi)$, we find $\alpha_c = 0.03 \pm 0.02$ as $\chi\rightarrow\infty$.
        Complex conjugation ($\hat{K}$) maps between skyrmion and antiskyrmion states, which leads to a crossing of the spectral lines and a degeneracy at $\alpha=0$.
        Panels~(b) and (c) show the energy spread $\braket{(\hat H - E)^2}/J^2$ at $\alpha=\{0,0.02,0.04\}$ versus the inverse bond dimension $1/\chi$ for skyrmions and antiskyrmions respectively.
    }
    \label{fig:cross_spread}
\end{figure*}

We consider a two-dimensional simple cubic lattice and a rectangular domain of $31\times15$ sites, motivated by the elongated shape of the skyrmion and antiskyrmion. 
It has been established that ground state properties of similar ordered magnetic materials can be reliably calculated via DMRG~\cite{Haller2022,Zhao2024,Bhowmick2024, Haller2024,Kuchkin2025_2}.
In this work, we find that excited states are well approximated by the DMRG-X algorithm~\cite{Khemani2016} [see~\Cref{fig:cross_spread}, panels~(b) and (c)].
We fix $J=1$ and $D=0.5$ so that the chosen system size is compatible with approximately one period of the spiral, given by $2\pi J/D\sim13$ in units of the interatomic distance, and to accommodate single-skyrmion excitations.
In the case of the classical monoaxial chiral magnet $(\alpha=0)$, the low-energy excitations are the saturated state, as well as spin-spiral, skyrmion, and antiskyrmion states. 
While the first two can be ground states, depending on the value of the uniform Zeeman field, the last two are always excited states.
To observe these states without additional surface twists on the system boundaries, we add boundary conditions in the form of classical environment spins, i.e., a single layer of classical spins around the simulation domain pinned along the $z$ axis~\cite{Siegl2022,Haller2024}.
Quantum spins interact with the classical environment through pairwise exchange and DM interactions similarly to \cref{eq:hamiltonian}, but where one spin operator is replaced by the pinned classical spin.
Therefore, the interaction with the classical environment is effectively an additional Zeeman interaction for the boundary quantum spins.
%

%%%%%%%%%%
\section{Results}
\label{sec:results}
%%%%%%%%%%

%%%%%
\subsection{Quantum skyrmions and antiskyrmions}
%%%%%

%
Skyrmion states can be found as excitations above the saturated state, for which we choose $B=0.6B_\mathrm{D}$. 
Excitations can be simulated via the DMRG-X method~\cite{Khemani2016}, which targets energy eigenstates that maximize fidelity to an initial state.
To target quantum (anti)skyrmion states, we start from classical (anti)skyrmion product states, more closely inspected in \cref{sec:prod_states}.
The standard DMRG-X algorithm scales with the bond dimension $\chi$ as $\chi^6$, and is therefore much more demanding than the standard DMRG algorithm used to approximate ground states, which instead scales with $\chi^3$~\cite{White1992}.
The energy spread $\braket{(\hat H - E)^2}$, evaluated for the converged DMRG-X states, signals that the quality of the approximation depends on the bond dimension $\chi$ [see \cref{fig:cross_spread}(b,c)] and reaches sufficient precision for $\chi_{\rm max} = 16$.
Skyrmion (antiskyrmion) states are numerically accessible for $\alpha > 0$ ($\alpha<0$), but convergence to antiskyrmion (skyrmion) excitations is found only in a narrow window near zero.
In the case $0<\alpha\leq1$, the skyrmion state $\ket{\psi_{\rm sk}}$ always has a lower energy than the antiskyrmion state $\ket{\psi_{\rm ask}}$ and vice versa for $-1 \leq \alpha < 0$ [see \cref{fig:cross_spread}(a)].
This necessarily leads to a level crossing of the two states at $\alpha = 0$, where the energies of the skyrmion and antiskyrmion states are equal.
The existence of a crossing can also be understood from the following symmetry arguments.
Consider the complex conjugation operator $\hat{K}$. 
The Hamiltonian is an operator-valued function of $\alpha$ satisfying $\hat{K}\hat{H}(\alpha)\hat{K}=\hat{H}(-\alpha)$.
Therefore, the anti-unitary operator $\hat K$ effectively inverts the sign of $\alpha$ in the Hamiltonian and maps the skyrmion eigenstates for $\alpha$ to antiskyrmion eigenstates for $-\alpha$.
At $\alpha = 0$, the Hamiltonian is real; therefore, it commutes with $\hat K$, which protects a twofold degeneracy.
The general state is an arbitrary superposition, i.e.
\begin{equation}
    \ket{\psi}
    =
    c \hat K
    \ket{\psi_{\rm sk}}
    +
    \sqrt{1-c^2}
    {\rm e}^{{\rm i}\varphi}
    \ket{\psi_{\rm sk}}
    .
    \label{eq:static_superposition}
\end{equation}
Spin expectation values of the symmetric superposition, $c=1/\sqrt2$, are shown in Fig.~\ref{fig:entropies}(b).
Since $\hat{K}\sigma_y\hat{K} = - \sigma_y $ (and $\sigma_x,\sigma_z$ are invariant), the conjugated state $\hat K\ket{\psi_{\rm sk}}$ has a magnetic texture in the spin expectation values that carries the negative topological charge of $\ket{\psi_{\rm sk}}$.
This justifies calling $\ket{\psi_{\rm ask}} = \hat K\ket{\psi_{\rm sk}}$ the corresponding antiskyrmion state.
Note that the skyrmion and conjugate skyrmion states are simultaneously eigenstates only at the $\hat K$-symmetric point $\alpha=0$.
%

%%%%%
\subsection{Product state approximation and entanglement}
\label{sec:prod_states}
%%%%%

\begin{figure}[h]
\centering
\includegraphics[width=\linewidth]{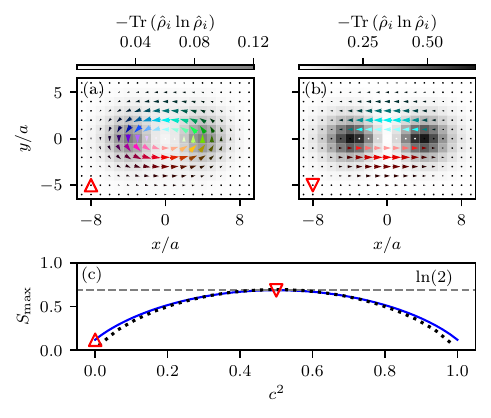}
\caption{
    Local von Neumann entanglement entropy presented as a grayscale background for the skyrmion state (a) and the symmetric superposition (b).
    The colored arrows represent the expectation values of local magnetization.
    Panel~(c) displays the maximum of the local von Neumann entropy for arbitrary superpositions.
    The computed $S_{\rm max}$ from MPS simulations (blue line) agrees very well with the analytic estimate in \cref{eq:Smax} (black dots).
    Note that due to the hermiticity of $\hat{\rho}_i$, its eigenvalues are invariant under complex conjugation, thus yielding the same entanglement entropy for both skyrmion and antiskyrmion states.
    }
\label{fig:entropies}
\end{figure}

%
In our work, we utilize product states either as initial states for the DMRG(-X) routines or as reference states to estimate the quantumness of the eigenstates.
We represent these states as products of spin-1/2 states with predetermined projections of the magnetic moment: 
\begin{align}
     \ket{\phi_{\rm sk}} = \prod_i \ket{\bm n_i} &=\prod_i e^{-{\rm i}\Phi_i \hat{S}_i^z} e^{-{\rm i}\Theta_i \hat{S}_i^y} \ket{\Uparrow}, \nonumber \\
      \braket{\phi_{\rm sk} | \hat{\bm S}_i | \phi_{\rm sk}} &= \braket{{\bm n}_i | \hat{\bm S}_i | {\bm n}_i} = \frac1{2}\bm n_i, 
      \label{ps}
\end{align}
with $\vert {\bm n}_i\vert=1$ for all $i$ and angles $\{\Phi_i,\Theta_i\}$ set to be local tilts of the magnetization arrows.
We choose the initial state of the DMRG-X algorithm as a product state with an arbitrary magnetic texture. For our purposes, this is typically the texture of a skyrmion.
In particular, we start from the product state $\prod_i\ket{\bm n(\Theta_i,\Phi_i)}$ (see \cref{sec:product_state_approx}) with the particular choice
\begin{align}
    \Theta_i &= 2 \arctan(\sinh( r_i / a)/\sinh(R/a)) - \pi
    ,\nonumber \\
    \Phi_i &= {\rm sign}(\alpha)\left[\phi(x_i,y_i) + \pi/2\right]
    ,\nonumber\\
    R &= 3a,
    \label{eq:DMRGX_initial}
\end{align}
and where $\phi$ denotes the azimuthal angle spanned by the $x$ and $y$ components of the spin.
Using bond dimension up to $\chi=16$ and applying DMRG-X, we achieve convergence to approximate eigenstates of $\hat H$, while keeping the bond dimension fixed to $\chi=1$ results in an energetically optimized product state.
For an antiskyrmion state, we have a product of complex-conjugated states:
\begin{equation}
     \ket{\phi_{\rm ask}} = \prod_i \hat{K} \ket{\bm n_i} \equiv \prod_i \ket{\overline{\bm n}_i},
     \label{ps_cc}
\end{equation}
where $\hat{K}$ is the complex conjugation operator.
The associated expectation values are  
\begin{equation}
      \braket{\phi_{\rm ask} | \hat{\bm S}_i | \phi_{\rm ask}} = \braket{\overline{\bm n}_i | \hat{\bm S}_i | \overline{\bm n}_i}= \braket{{\bm n}_i | \hat{\bm S}^*_i | {\bm n}_i} = \frac1{2}\overline{\bm n}_i
      ,
\end{equation}
and it is straightforward to recognize that the magnetization profiles of the skyrmion and antiskyrmion product states are related via a sign exchange of the $y$ component, $(\overline{n}_{i,x},\overline{n}_{i,y},\overline{n}_{i,z})=(n_{i,x}, -n_{i,y},n_{i,z})$.
While individual product states lack the key quantum features of an MPS, such as entanglement, they can be used to predict the qualitative behavior of individual (anti)skyrmion states, thus serving as an adequate approximation in a handful of cases. 
Several measures may be introduced to quantify to which extent a semi-classical product state acceptably approximates an MPS.
One such quantity is the local von Neumann entanglement entropy.
To calculate the von Neumann entropy at site $i$, we construct the reduced density matrix,
\begin{equation}
    \hat\rho_i = {\rm Tr}_{\bar i}\ket{\psi}\bra{\psi}
\end{equation}
where ${\rm Tr}_{\bar i}$ denotes a partial trace over the entire complement of $i$ and $\ket{\psi}$ represents the MPS. 
The entropy is then defined as:
\begin{equation}
    S_i = - {\rm Tr}(\hat{\rho}_i \ln{\hat{\rho}_i}).
\end{equation}
If the site $i$ is completely disintangled from its environment, as is the case for product states, its reduced matrix becomes a projector with eigenvalues $\{ 0,1 \}$, leading to vanishing entropy.
In Fig.~\ref{fig:entropies}, we present the von Neumann entropy calculated for different matrix product states with bond dimension $\chi=16$.
We observe that quantum effects play a subleading role for individual (anti)skyrmion states.
This manifests in relatively low local entanglement compared to the maximal value ($\ln 2$ for spin-$1/2$).
Thus, we conclude that the product states with sharp local magnetization are an adequate candidate to approximate the (anti)skyrmion MPS states.
In contrast, the entanglement increases steeply at two regions of non-collinearity of the superposition state in Fig.~\ref{fig:entropies}(c), which signals large deviations from a product state.
Within these regions, the expected local magnetization also tends to zero.
These features can be explained by a coherent superposition of skyrmion-antiskyrmion product states, i.e., by the approximation $\ket\psi\approx c^2\ket{\phi_{\rm ask}} + \sqrt{1-c^2}{\rm e}^{\rm i\varphi}\ket{\phi_{\rm sk}}$.
At sites $i'$ where $\braket{\phi_{\rm ask} | \hat S_{i',z} | \phi_{\rm ask}} = 1/2$, the reduced density matrix is given by $\hat\rho_{i'} = \ket\uparrow\bra\uparrow$, leading to vanishing entanglement.
In contrast, due to the conjugate nature of the (anti)skyrmion states, at sites where the $y$ component of the spin expectation value reaches its maximum value, we find $\hat\rho_{\rm max} \approx c^2\ket+\bra+ + (1-c^2)\ket-\bra-$, where $\ket\pm$ denote the eigenstates of the spin operator along the $y$ direction (we neglect vanishingly small overlap elements which we discuss in more detail in \cref{sec:orthogonality_centers}).
This allows us to estimate the entropy in the superposition as 
\begin{align}
    S_{\rm max} \approx -c^2\ln c^2 - (1-c^2)\ln(1-c^2)
    ,
    \label{eq:Smax}
\end{align}
which is in very good agreement with the simulated results presented in \cref{fig:entropies}(c).

It is straightforward to evaluate the spin polarization at these points for generic superpositions, i.e., $n_{y} = 2{\rm Tr}{\hat S_{{\rm max},y}\hat\rho_{\rm max}} \approx 2c^2-1$, such that we establish a connection between entanglement entropy and average magnetization
\begin{align}
    S_{\rm max} \approx \frac12 \left(\ln4-\ln(1-n_y^2) - n_y\ln\left(\frac{1+n_{y}}{1-n_{y}}\right)\right)
    .
    \label{eq:Smax2}
\end{align}
In conclusion, for symmetric superpositions $c^2=1/2$, the local magnetization is strongly suppressed, and spins located at these special points are maximally entangled with the environment [see \cref{fig:entropies}(b)].

%%%%%
\subsection{Orthogonality centers}
\label{sec:orthogonality_centers}
%%%%%

%
The previous section concerned the appearance of two regions with reduced magnetization length and increased entanglement for $0<c<1$.
In the product state approximation, it can be shown that the centers of these regions are tied to the orthogonality between the magnetization profiles of skyrmion and antiskyrmion.
The mutual overlap of the states $\ket{\phi_{\rm sk}}$ and $\ket{\phi_{\rm ask}}$ can be decomposed into a product of local overlaps for each individual node,
\begin{align}
\braket{\phi_\mathrm{sk}|\phi_\mathrm{ask}} &= \prod_i\braket{{\bm n}_i | \overline{\bm n}_i} \equiv \prod_i v_i \nonumber \\
& = \prod_i \left(\cos\Phi_{i}+\mathrm{i}\sin\Phi_{i}\cos\Theta_{i}\right), 
\label{eq:overlap}
\end{align}
where the parametrization
\begin{align}
  \ket{\bm{n}_{i}}
  &=
  \ket{\bm{n}(\Theta_i,\Phi_i)} \nonumber\\
  &=\left(e^{-\mathrm{i}\Phi_{i}/2}\cos\Theta_{i}/2, e^{\mathrm{i}\Phi_{i}/2}\sin\Theta_{i}/2\right)^\mathrm{T} 
\end{align}
is used.
Then, one finds the magnetization vectors $\bm{n}(\Theta_i,\Phi_i)=\left(\sin\Theta_{i}\cos\Phi_{i},\sin\Theta_{i}\sin\Phi_{i},\cos\Theta_{i}\right)$, so that one obtains a phase-gauge independent expression,
\begin{equation}
    |v_{i}|^{2}=\left(1-n_{i,y}\right)\left(1+n_{i,y}\right).
\end{equation}
Therefore, at sites $i$ where the $y$-component of spin expectation values reaches $\pm1$, the overlap elements $v_i$ vanish.
As long as this condition is met anywhere, the inner product \eqref{eq:overlap} is dominated by a zero overlap and vanishes as well.
For this reason, we call these special points orthogonality centers.

Due to the spatial separation of the orthogonality centers, energy-degenerate skyrmions and antiskyrmions couple only weakly via local interactions.
This can be recognized in the matrix elements of the spin operator, which are strongly suppressed by the product of local overlap elements $v_i$, i.e.
\begin{align}
    \braket{\psi_{\rm sk} | \hat{\bm S_i} | \psi_{\rm ask}}
    &=
    \braket{{\bm n}_i | \hat{\bm S_i} | \overline{{\bm n}}_i} \prod_{j\neq i} v_j, \nonumber \\
    \braket{{\bm n}_i | \hat{\bm S_i} | \overline{{\bm n}}_i}
    &=
    \dfrac{1}{2}\begin{pmatrix}
        \sin\Theta_{i}
        \\
        0
        \\
        \cos\Theta_i\cos\Phi_i + {\rm i}\sin\Phi_i
    \end{pmatrix}
    .
    \label{eq:prod_spin_exp}
\end{align}
Note that in Eqs.~\eqref{eq:overlap} and \eqref{eq:prod_spin_exp}, the last equations are phase-gauge dependent.
For the parameters used, we find that the local orthogonality is not perfect, which allows us to exploit a direct coupling mechanism:
Consider a perturbation of the form $\delta \hat H = \delta B_z \hat S_z$, where $\hat S_z=\sum_i\hat S_{i,z}$ is the total $z$-magnetization operator.
In the two-dimensional subspace $\ket{\bm\Psi} = (\ket{\psi_{\rm sk}},\ket{\psi_{\rm ask}})$, this corresponds to the effective Hamiltonian
\begin{align}
    & \delta \hat H_{\rm eff}
    =
    \ket{\bm\Psi}
    \begin{pmatrix}
        a & |b| {\rm e}^{-{\rm i}\varphi} \\
        |b| {\rm e}^{{\rm i}\varphi} & a
    \end{pmatrix}
    \bra{\bm\Psi}, \\
    &
    a = \delta B_z\braket{\psi_{\rm sk} | \hat S_z | \psi_{\rm sk}}
    ,\ 
    b = \delta B_z\braket{\psi_{\rm sk} | \hat S_z | \psi_{\rm ask}} \notag 
    .
\end{align}
The diagonal element $a$ is proportional to the total $z$-magnetization of the (anti)skyrmion state, while the off-diagonal matrix element $b$ is small.
The eigenvalues of $\delta \hat H_{\rm eff}$ are $a\pm |b|$ and two orthonormal eigenstates are given by $(\ket{\psi_{\rm sk}}\pm{\rm e}^{{\rm i}\varphi}\ket{\psi_{\rm ask}})/\sqrt2$.
These correspond to the symmetric $c=1/\sqrt2$ superposition states of~\cref{eq:static_superposition} and have the intriguing spin expectation value profile shown in~\cref{fig:entropies}(b).
In the continuum limit of the classical mean-field theory, there necessarily exist two points where $\bm n_{\rm sk} = \pm  \mathbf{e}_y$, and therefore $b$ vanishes, which highlights the importance of the lattice and the small size of the (anti)skyrmions.
With sufficient experimental control over $\alpha$ and $\delta B_z$, the effective total Hamiltonian of the degenerate subspace could be used to realize the $x$-rotation and phase shift gates in skyrmion-based quantum computation~\cite{Psaroudaki2023,Xia2023,Chudnovsky2025}.
This (anti)skyrmion charge approach to skyrmion qubits could be particularly fruitful because of their energetic separation from other states.
Physically, the $c=1/\sqrt2$ superposition states are the magnetic analogs of mesoscopic Schrödinger cat states.
Under the influence of a magnetic field gradient, the skyrmion and antiskyrmion states have a different time evolution as predicted by Thiele's equation~\cite{Thiele_73} applied to our particular setup, see~\cref{sec:thiele}.
For the superposition state, this leads to a spatial separation of the skyrmion and antiskyrmion wave functions over time, which we will discuss further below (see~\cref{fig:fig1}).
%

%%%%%
\subsection{Magnetic SANS as a probe of skyrmionic Schr\"odinger cat states}
%%%%%

%
The properties of the skyrmion state $\ket\psi$ may be probed using the magnetic small-angle neutron scattering (SANS) technique (see, e.g., Refs.~\cite{rmp2019,dirkreview2022} for reviews).
This method is a key experimental tool, since it offers, quite uniquely, the possibility to investigate the magnetic microstructure in the volume of magnetic media and on the relevant mesoscopic length scale ($\sim 1-1000~{\rm nm}$).
The scattering contrast for magnetic SANS is related to spatial nanometer-scale variations in the magnitude and orientation of the magnetization texture.
Here, we calculate a set of SANS observables for the elastic scattering of a neutron beam polarized parallel to the field ${\bm B}=B{\bm e}_z$ with a primary focus on the so-called spin-flip cross section.
The latter quantity can be measured using uniaxial polarization analysis and is given by
\begin{equation}
    \frac{{d}\Sigma_{\rm sf}}{{d}\Omega}
    \propto
    \sum_{\alpha,\beta=x,y}
    \left(
        \delta_{\alpha\beta}
        -
        \frac{q_\alpha q_\beta}{q^2}
    \right)
    \mathcal{S}_{\alpha\beta}({\bm q}),
    \label{eq:spin_flip_cross_section}
\end{equation}
where ${\bm q}=(q_x,q_y,q_z)$ is the momentum transfer vector and $\mathcal{S}_{\alpha\beta}({\bm q})$ are the structure factors
\begin{equation}
  \mathcal{S}_{\alpha\beta}(\bm q)
  =
  \sum_{ij} e^{-i\bm q \cdot(\bm r_i - \bm r_j)}
  \braket{\psi | \hat S _{i,\alpha} \hat S_{j,\beta} | \psi}.
  \label{eq:structure_factors}
\end{equation}
The unpolarized magnetic SANS cross section $d\Sigma/d\Omega$ is similar to \cref{eq:spin_flip_cross_section}, but the sum is taken over all components $\{x,y,z\}$ of the structure factors.
In the small-angle regime, the component of $\bm{q}$ along the incident beam is significantly smaller than the other two components, i.e. $q_z \approx 0$.
This implies that the only difference between the spin-flip and the unpolarized SANS cross section is a single structure factor, $d\Sigma/ d\Omega-d\Sigma_{\rm sf}/d\Omega=\mathcal{S}_{zz}$.
In the product state approximation (see~\cref{sec:prod_states,sec:orthogonality_centers}), $\mathcal{S}_{zz}$ forms a common background for skyrmions and antiskyrmions. 
The background variation is largely suppressed for $0 < c < 1$ due to the opposite local order between the skyrmion and antiskyrmion states that contribute to the superposition.
For this reason, $\mathcal{S}_{zz}$ effectively represents the same shift for all amplitudes $c$, so individual superpositions could also be detected in unpolarized SANS experiments.
\begin{figure}[h]
    \centering
    \includegraphics[width=\columnwidth]{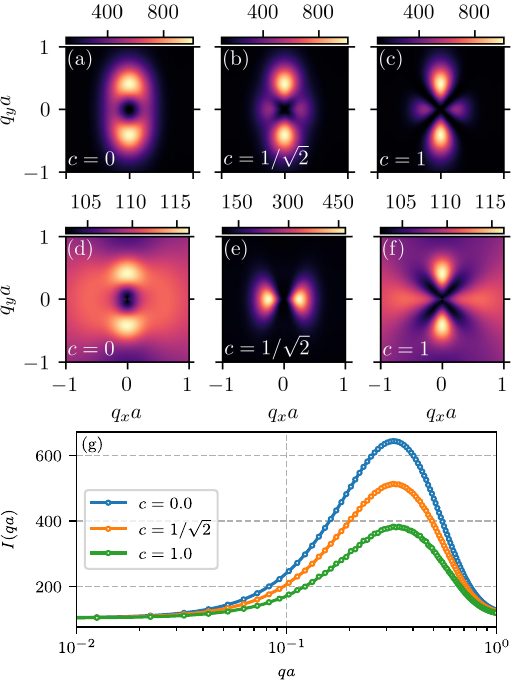}
    \caption{
        Panels (a)$-$(c) display the spin-flip SANS cross sections ${d}\Sigma_{\rm sf}/{d}\Omega$ in arbitrary units for skyrmion state ($c=0$), symmetric superposition ($c=1/\sqrt{2}$), and antiskyrmion state ($c=1$).
        Panels (d)$-$(f) show their connected parts. 
        Note the change of scale in (e) compared to (d) and (f).
        The signature of the superposition state manifests in two pronounced peaks along $q_x$.
        (g)~Azimuthally-averaged spin-flip SANS intensity [of the data from (a)$-$(c)] as a function of $qa$ (log-linear scale).
    }
    \label{fig:sf_sans}
\end{figure}

In \cref{fig:sf_sans}(a)$-$(c) we display the spin-flip SANS cross section for different superpositions.
The equal-part superposition bears qualitative aspects of both the skyrmion and antiskyrmion cross sections, i.e. ${d\Sigma_\mathrm{sf}}/{d\Omega} \approx c^2{d\Sigma_\mathrm{sf}^{\rm ask}}/{d\Omega} + (1-c^2){d\Sigma_\mathrm{sf}^{\rm sk}}/{d\Omega}$. 
This can also be observed in the spin-flip intensity $I_{\rm sf}(q)$ obtained by averaging out the azimuthal degree of freedom in the $q_x$-$q_y$~plane
\begin{equation}
    I_{\rm sf}(q) = \frac{1}{2\pi}\int_0^{2\pi} \frac{d\Sigma_{\rm sf}}{d\Omega}(q,\theta)d\theta .
\end{equation}
The results for $I_{\rm sf}(q)$ are dsiplayed in \cref{fig:sf_sans}(g), revealing that the signal of the symmetric superposition corresponds to the average between the skyrmion and antiskyrmion cases. 
In contrast, the differences between semi-classical and superposition states are well pronounced in the {\it connected} part of the spin-flip cross section ${d}\Sigma_\mathrm{sf}^{\rm conn}/{d}\Omega$, where in \cref{eq:structure_factors} we replace 
\begin{equation}\label{eq:conn}
    \begin{split}
    \braket{\hat S_{i,\alpha} \hat S_{j,\beta}} 
    &\rightarrow 
    \langle\!\langle \hat S_{i,\alpha} \hat S_{j,\beta} \rangle\!\rangle \\
    \langle\!\langle \hat S_{i,\alpha} \hat S_{j,\beta} \rangle\!\rangle 
    &= 
    \braket{ \hat S_{i,\alpha} \hat S_{j,\beta}} - \braket{\hat S_{i,\alpha}}\braket{\hat S_{j,\beta}}.
    \end{split}
\end{equation}
\Cref{fig:sf_sans}(d)$-$(f) shows that this observable is ideally suited to detect the difference between states with high and zero site-to-site entanglement.
As we show in \cref{sec:prod_states} and \cref{sec:product_state_approx}, (anti)skyrmion matrix product states, obtained from the DMRG-X algorithm, exhibit significant overlap with simple product state approximations.
However, these product-state approximations yield no special structure in the connected cross sections.
Therefore, the structure of the signal in \cref{fig:sf_sans}(d) and \cref{fig:sf_sans}(f) emerges exclusively from quantum effects.
Note that the variations between the minima and maxima in \cref{fig:sf_sans}(d) and \cref{fig:sf_sans}(f) are approximately a factor $20$ smaller compared to \cref{fig:sf_sans}(e) for the chosen parameters.
The pronounced double-peak structure can instead be attributed to the state superposition alone (see~\cref{sec:product_state_sans}, \cref{fig:ps_sans}).
%

%%%%%
\subsection{Dynamics under a magnetic field gradient}
%%%%%

\begin{figure*}[ht!]
    \centering
    \includegraphics[width=\linewidth]{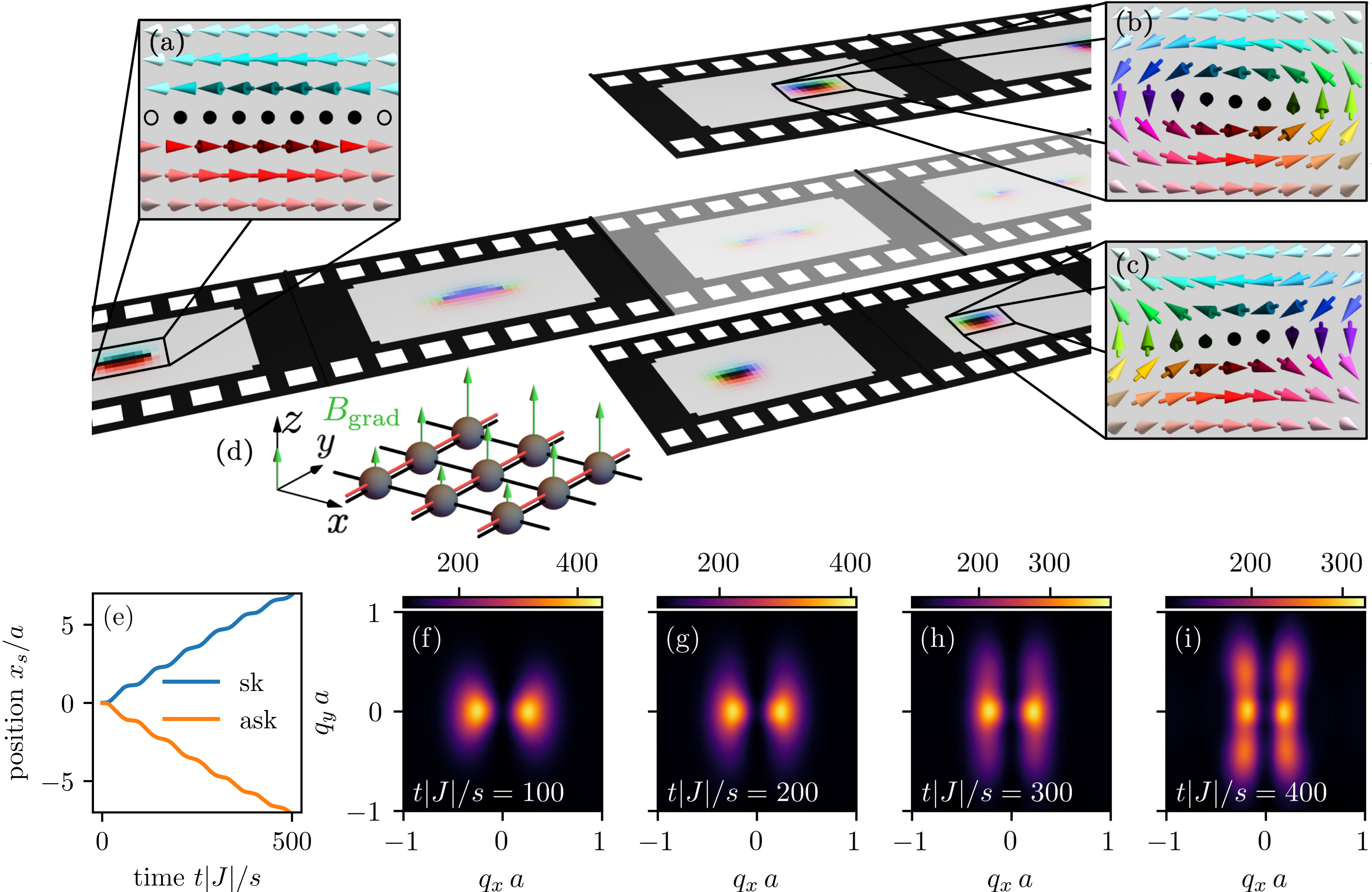}
    \caption{
        The gradient-field-induced time evolution of the Schrödinger cat state shown in panel (a), which is an equal superposition of a symmetric skyrmion (b) and antiskyrmion (c).
        A projective measurement made after the second snapshot (illustrated by transparency) leads to the collapse of the wave function, and, as a result, the system evolves as a classical skyrmion or antiskyrmion.
        Panel~(d) shows a sketch of the square lattice, where the nodes represent spin-$1/2$ sites.
        Black and red bonds correspond to exchange and Dzyaloshinskii-Moriya interactions, respectively, and green arrows show the Zeeman field perturbation.
        Panel~(e) displays the change of the average position for (anti)skyrmions under the field $B_{\rm grad}/J = -0.02$.
        Panels~(f)$-$(i) display the connected part of the spin-flip SANS cross section at times $t|J|/s\in\{100,200,300,400\}$, calculated for the symmetric superposition state.
    }
    \label{fig:fig1}
\end{figure*}

%
We consider the dynamics of (anti)skyrmion excitations under the influence of a magnetic field with a small gradient along the $y$ direction.
For this purpose, we numerically solve the Schrödinger equation ${\rm i}\partial_t\ket\psi = \hat H\ket\psi$ using the time-dependent variational principle (TDVP)~\cite{Haegeman2016,Paeckel2019} (see Appendix~\ref{sec:tdvp} for convergence results).
The superposition state of Eq.~\eqref{eq:static_superposition} evolves according to
\begin{align}
    \ket{\psi(t)}
    =
    c
    \ket{\psi_{\rm ask}(t)}
    +
    \sqrt{1-c^2}{\rm e}^{{\rm i}\varphi}\ket{ 
    \psi_{\rm sk}(t)}
    .
    \label{Superposition}
\end{align}
Due to the $\hat K$-symmetry of $\hat H$ at $\alpha=0$, the trajectories of the initial states are given by
\begin{align}
    \ket{\psi_{\rm sk}(t)}
    &=
    {\rm e}^{-{\rm i}\hat Ht}
    \ket{\psi_{\rm sk}(0)}
    ,
    \\
    \ket{\psi_{\rm ask}(t)}
    &=
    {\rm e}^{-{\rm i}\hat Ht}
    \hat K\ket{\psi_{\rm sk}(0)}
    =
    \hat K
    \ket{\psi_{\rm sk}(-t)}
    \label{eq:ask_evolution}
    .
\end{align}
Therefore, skyrmions and antiskyrmions share a particle-antiparticle relationship: they have opposite topological charges, and their dynamics are related via time reversal.
To demonstrate a coherent motion of the wave, we compute spin expectation values at times $t$,
\begin{align}
    \bm m_{i,s}(t)
    &=
    \braket{\psi_{s}(t) | \hat{\bm S}_i | \psi_{s}(t)}
    ,
    \
    s\in\{\rm sk, ask\}
    ,
\end{align}
and analyze the evolution of the average skyrmion position, $(x_{s}(t),y_{s}(t)) = \sum_i \bm{r}_i m_{i,s}^z(t) / \sum_i m_{i,s}^z(t) $ (see~\cref{fig:fig1}).
The skyrmion and antiskyrmion move with opposite Hall angles, which is compatible with the solutions of Thiele's classical equation for rigid skyrmion motion in the absence of dissipation, i.e., $\dot{x}_{s}\propto B_{\rm grad}/Q$, $\dot{y}_s=0$, where $Q=\pm1$ is the topological charge (see Appendix~\ref{sec:thiele}). 
In \cref{fig:fig1}(e), we display the change of the average position for (anti)skyrmions under the Zeeman field perturbations.
At the perturbation amplitude $B_{\rm grad}/J = -0.02$,
we find an average velocity of $v_s \approx 0.014 |J|Qa$ via a linear fit of the trajectories.
This presents an adequate match to the result $v_s\approx 0.0137 |J|Qa$ obtained from the Thiele equation.
Similarly to \cref{sec:prod_states}, we approximate $\ket{\psi(t)}\approx c\ket{\phi_{\rm ask}(t)} + \sqrt{1-c^2}{\rm e}^{\rm i\varphi}\ket{\phi_{\rm sk}(t)}$, and find
\begin{align}
    \braket{\psi(t) | \hat{\bm S}_i | \psi(t)}
    \approx
    c^2\bm m_{i,\rm ask}(t)
    +
    (1-c^2)\bm m_{i,\rm sk}(t)\label{sk_ask_dyn}
    ,
\end{align}
where we again neglect small matrix elements $\braket{\phi_{\rm sk}(t) | \hat{\bm S}_i | \phi_{\rm ask}(t)}$.
Note that the matrix elements become more suppressed over time as the skyrmion and antiskyrmion centers move in opposite directions.
In conclusion, we find that the spin expectation value equals the weighted average, which is further confirmed by the MPS simulations.
This dynamics can be qualitatively described with the classical Landau-Lifshitz equation by simulating the (classical) skyrmion and antiskyrmion wavefront separately and then constructing the superposition state of interest according to~\cref{sk_ask_dyn}.
Due to subleading quantum effects in the individual wavefronts, the resulting deviation is minor (see the comparison in~\cref{fig:tdvp_different_M}).
In an experiment, the spin expectation value would result from projective measurements such that the superposition state collapses to either skyrmion or antiskyrmion states.
Consequently, the hypothetical observer would measure $\bm m_{i,\rm ask}(t)$ or $\bm m_{i,\rm sk}(t)$ with probability $c^2$ or $1-c^2$, respectively.
To further characterize the wavefront behavior, we examine the connected part of the polarized SANS cross section at different times [see \cref{fig:fig1}(f)$-$(i)].
For the symmetric superposition state, we observe a broadening of the two central peaks along $q_y$, indicative of distinct wavefront dynamics in the symmetric superposition states.
This behavior can be understood from two features of the structure factor components that we derive in~\cref{sec:product_state_sans}, i.e., an increasing signal $\propto\sin^2(\bm q\cdot \bm vt)\mathcal F_{x}(\bm q)$ with four peaks around the origin, and a decreasing component $\propto\cos^2(\bm q\cdot\bm v t)\mathcal F_{y}(\bm q)$ with two peaks around the origin.
Tracking the temporal change of the overall amplitude of these peaks thus allows to estimate the velocity of the wavefront.
%

%%%%%%%%%%
\section{Conclusions}
\label{summary}
\vspace{-0.4cm}  % squeeze a bit to fit on the last page
%%%%%%%%%%

%
In this work, we studied the static and dynamic properties of quantum skyrmion and antiskyrmion excitations in monoaxial chiral magnets. 
Using symmetry arguments and density matrix renormalization group simulations, we demonstrated that these states become energetically degenerate in the monoaxial limit, allowing the formation of magnetic Schrödinger cat states.
We explored how the quantum skyrmion, antiskyrmion, and symmetric superposition states can be characterized through spin-spin correlation functions and their signatures in polarized neutron scattering.
In particular, we found that the individual states display qualitative distinctions in theoretically calculated spin-flip SANS cross sections--which could serve as fingerprint observables for future experiments.
Additionally, we investigate the coherent time evolution of the superposition state under a magnetic field gradient.
Our results show spatial separation of the skyrmion and antiskyrmion components, in close agreement with semi-classical predictions of Thiele's equation. 
Taken together, our findings establish a theoretical framework for the detection, manipulation, and potential control of skyrmionic Schrödinger cat states.
They suggest that sufficient control over the DMI anisotropy and the gradient field can open up new avenues for quantum computation using quantum skyrmions as qubits.

%%%%%%%%%%
\vspace{-0.5cm}  % squeeze a bit to fit on the last page
\section*{Acknowledgements}
\vspace{-0.4cm}  % squeeze a bit to fit on the last page
%%%%%%%%%%

We acknowledge financial support from the National Research Fund of Luxembourg under Grants C22/MS/17415246/DeQuSky and AFR/23/17951349. 
V.M.K. acknowledges the financial support from the European Union’s Horizon 2020 research and innovation programme under the Marie Sk{\l}odowska-Curie grant agreement No.~101203692 (QUANTHOPF).

%%%%%%%%%%
\vspace{-0.5cm}  % squeeze a bit to fit on the last page
\section*{Data Availability}
\vspace{-0.4cm}  % squeeze a bit to fit on the last page
%%%%%%%%%%
The raw data is available upon reasonable request.

\bibliographystyle{apsrev4-2}
\bibliography{bib_submission.bib}

\newpage
\clearpage

\appendix

%%%%%%%%%%
\section{Local fidelity}
\label{sec:product_state_approx}
%%%%%%%%%%

\begin{figure*}[ht!]
    \centering
    \includegraphics{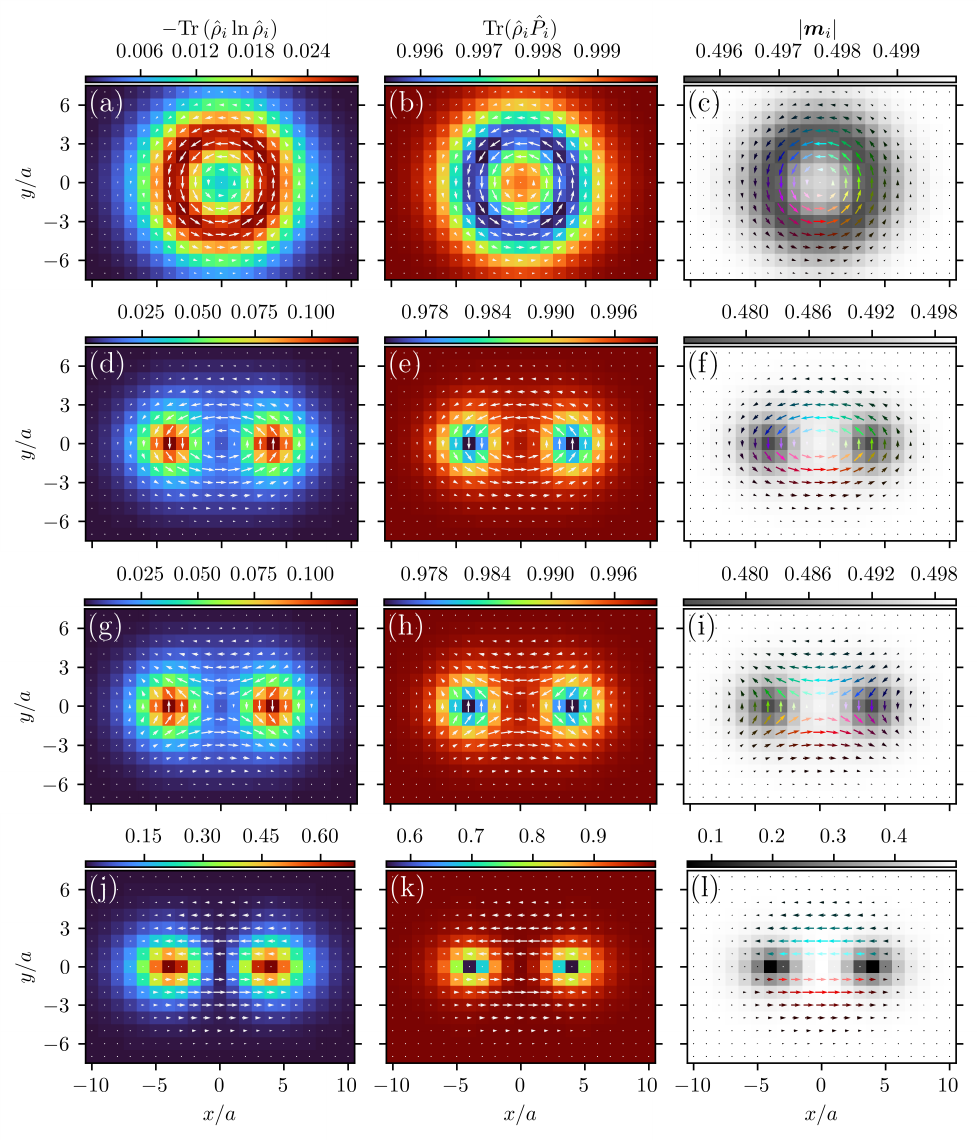}
    \caption{Panels (a)$-$(c) show the spatial profiles of the von-Neumann entropy, the local fidelity and the magnetization norm, respectively, calculated for a skyrmion MPS obtained for an axially symmetric magnet $(\alpha=1
    )$. Panels (d)$-$(f) and (g)$-$(i) display, respectively, the same local observables for a skyrmion and antiskyrmion MPS in a monoaxial magnet. The last row (k)$-$(l) corresponds to the symmetric superposition of the skyrmion and antiskyrmion states. The used DMI strength is $D=J/2$ with $B=0.8B_\mathrm{D}$ for the symmetric skyrmion and $B=0.6B_\mathrm{D}$ for the monoaxial states.}
    \label{fig:fidelities}
\end{figure*}

%
In addition to the local von Neumann entropy, one may introduce a number of observables to gauge the quantum nature of an MPS. 
To measure how closely a product state resembles an arbitrary MPS, we calculate the local fidelity, effectively quantifying local overlaps between a product state and an MPS. 
In order to do so, we construct the local projector onto the Pauli basis:
\begin{equation}
    \hat P(\bm m_i) = \hat P_i = \frac{1}{2}\left(\mathds 1 + \frac{\bm m_i \cdot \bm \sigma_i}{\vert \bm m_i \vert}\right),
\end{equation}
where $\bm m_i = \braket{\psi \vert \hat{\bm S}_i \vert \psi}$, $\ket{\psi}$ is the MPS and $\bm \sigma_i$ is the matrix of Pauli matrices on site $i$.
We define local fidelity as the expectation value of $\hat P_i$, i.e., $\mathrm{Tr}(\hat\rho_i\hat P_i)$, where $\hat{\rho}_i = {\rm Tr}_{\bar i}\ket{\psi}\bra{\psi}$ is the reduced density matrix.
We present the local fidelity, the von-Neumann entropy $S_i = -\mathrm{Tr} (\hat{\rho}_i\ln\hat{\rho}_i)$ and the magnetization norm $\vert \bm m_i \vert$ in \cref{fig:fidelities}.
These local observables are calculated for monoaxial skyrmion and antiskyrmion MPS wave functions, their symmetric superposition, and a skyrmion MPS for a system with isotropic DMI ($\alpha=1$).
We observe that the entropy is bound to regions of non-collinear spins.
For the isotropic DMI case in Fig.~\ref{fig:fidelities}(a)$-$(c), this region happens to be located at the perimeter of the axially symmetric skyrmion where the magnetization arrows lie in the $xy$ plane.
For the monoaxial DMI case, these regions are deformed into two spatially separate domains which accommodate the elliptical shape of the skyrmion texture. 
The suppressed fidelity regions coincide with regions with an increase in von-Neumann entanglement entropy due to quantum fluctuations and regions with a reduced magnetization norm.
This supports the statement that these observables, though not strictly equivalent, consistently indicate quantum behavior in a similar fashion. 
%

%%%%%%%%%%
\section{SANS cross sections of product states}
\label{sec:product_state_sans}
%%%%%%%%%%

%
Calculating the SANS observables of product states allows us to gauge the neutron response of our matrix product states against responses of states that behave classically.
We will proceed to show that a symmetric superposition of product states produces qualitatively the same signal as the MPS analogue in a scattering observable that we dub the (un)polarized connected cross section:
\begin{equation}
    \frac{d\Sigma^{\rm C}}{d\Omega} = \sum_{\alpha\beta}\left(\delta_{\alpha\beta}-\frac{q_\alpha q_\beta}{q^2}\right)\mathcal{S}^{\rm C}_{\alpha\beta},
    \label{eq:cross section}
\end{equation}
where $\alpha,\beta\in\{x,y\}$ for the spin-flip SANS cross section, $\alpha,\beta\in\{x,y,z\}$ for the unpolarized cross section, and $q_z\approx 0$.
In the calculation of the structure factors $\mathcal{S}^{\rm C}_{\alpha\beta}$, the calculation of two-point correlation functions becomes relevant. 
The matrix elements are given by
\begin{align}
    \braket{\phi_{\rm sk} | \hat{\bm S_i} \hat{\bm S}_j | \phi_{\rm ask}}
    =
    \braket{{\bm n}_i | \hat{\bm S_i} | \overline{{\bm n}}_i} \braket{{\bm n}_j | \hat{\bm S}_j | \overline{{\bm n}}_j} \prod_{k\neq i,j} v_k
    ,
\end{align}
and are also largely suppressed by the overlap elements $v_k$ discussed in \cref{sec:orthogonality_centers}.

The structure factors for the connected cross section are defined as the Fourier transforms of the connected correlation $\Gamma_{ij}^{\alpha\beta} \equiv \braket{ \hat{S}_{i,\alpha} \hat{S}_{j,\beta}}-\braket{\hat{S}_{i,\alpha}  }\braket{\hat{S}_{j,\beta}}$:
\begin{equation}
    \mathcal{S}^{\rm C}_{\alpha\beta}
    =
    \sum_{ij}
    e^{-{\rm i}{\bm q}\cdot({\bm r}_i-{\bm r}_j)} \Gamma_{ij}^{\alpha\beta},
    \label{eq:structure}
\end{equation}
where the wave function is, in general, the superposition of the skyrmion and antiskyrmion product states, $\ket{\phi}=c\ket{\phi_{\rm ask}}+\sqrt{1-c^2}e^{{\rm i}\varphi}\ket{\phi_{\rm sk}}$.
The diagonal elements $i=j$ in the transform yield a $\bm q$-independent contribution and therefore shall be ignored for now, as they only form a constant background on the level of the structure factors.
The individual off-diagonal components read as
\begin{align}
    \Gamma_{ij}^{\alpha\beta}
    &=
    \frac1{4}c^2(1-c^2)
    \left(
        n_{i,\alpha}n_{j,\beta}
        +
        \overline{n}_{i,\alpha}\overline{n}_{j,\beta}
    \right.
    \nonumber \\
    & \qquad
    \left.
        - n_{i,\alpha}\overline{n}_{j,\beta} - \overline{n}_{i,\alpha}n_{j,\beta}
    \right)
    ,
\end{align}
where terms of the order $\prod_{k\neq i,j}\vert v_k \vert$ and smaller are neglected.
It follows that in the case of exactly overlapping textures, the only non-vanishing contribution corresponds to $\alpha=\beta=y$ and reaches its maximum at $c^2=1/2$, which yields $\Gamma_{ij}^{yy} = \frac1{4}n_{i,y}n_{j,y}$.
The Fourier transform of this connected part is solely responsible for the distinct Bragg peaks in the SANS signal of the equal-part superposition.
In the case of displaced skyrmion and antiskyrmion textures, other structure factors become prominent.
Under the influence of the magnetic field gradient $\bm B^{\rm grad}$ the textures drift apart in opposite directions.
The velocity $\bm v$ relative to the origin is calculated in Eq.~\eqref{eq:velocity} for the continuum limit.
Adopting a reference frame connected with the skyrmion texture, we write the Fourier components in the continuum limit as follows:
\begin{align}
    &
    \Gamma^{\alpha\beta}({\bm r},{\bm r}') 
    =
    \frac{1}{4}c^2(1-c^2)\big[n_\alpha({\bm r}) n_\beta({\bm r}')
    \nonumber\\
    &\quad+
    \overline{n}_\alpha({\bm r}) \overline{n}_\beta({\bm r}' - n_\alpha)({\bm r}) \overline{n}_\beta({\bm r}') 
    -\overline{n}_\alpha({\bm r}) n_\beta({\bm r}')\big].
\end{align}
Using the relation $\overline{ {\bm n}}({\bm r})= {\rm diag}(1,-1,1){\bm n}({\bm r}-2{\bm v}t)$, we find that the structure factors inherit a time dependence
\begin{align}
    \mathcal{S}^{\rm C}_{yy}
    &=
    c^2(1-c^2)\cos^2({\bm q}\cdot{\bm v} t) \mathcal{F}_{yy}(\bm q), \notag \\
    \mathcal{F}_{\alpha\beta}(\bm q)
    &=
    \int d^3rd^3r' e^{-{\rm i}{\bm q}\cdot ({\bm r}-{\bm r}')}n_\alpha(\bm r)n_\beta(\bm r').
\end{align}
The structure factor above is predominant as long as $t\ll(\bm q \cdot \bm v)^{-1} $.
Analogously, $\mathcal{S}^{\rm C}_{xx}$ and $\mathcal{S}^{\rm C}_{zz}$ grow parabolically on this time scale, i.e.,
\begin{equation}
    \mathcal{S}^{\rm C}_{\alpha\alpha}
    =
    c^2(1-c^2)\sin^2({\bm q}\cdot{\bm v} t) \mathcal{F}_{\alpha\alpha}(\bm q),
\end{equation}
for $\alpha \in \{x,z\}$.

\clearpage

\begin{figure}[ht]
    \centering
    \includegraphics[width=\columnwidth]{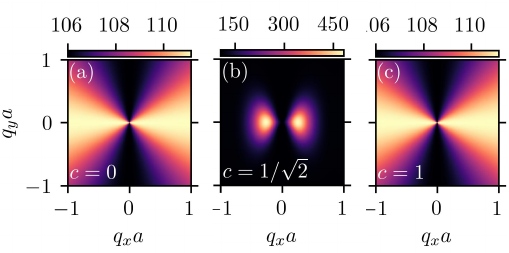}
    \caption{
        The connected cross section $d\Sigma^{\rm C}/d\Omega$ of the skyrmion product state (a), antiskyrmion product state (c) and their symmetric superposition (b).
        Due to the relative sign flip of $n_{i,y}$ between the skyrmion and antiskyrmion textures, their SANS signals obtain opposite corrections in the subleading order.
        Since the amplitude of the corrections is small, panels (a) and (b) appear to be identical.
    }
    \label{fig:ps_sans}
\end{figure}

For the product states ($c\in \{0,1\}$), the Fourier components $\Gamma_{ij}^{\alpha\beta}$ vanish under the assumption that $i\neq j$.
In this case, the only contribution stems from the diagonal elements related to on-site correlations:
\begin{align}
    \mathcal{S}^{\rm C}_{\alpha\beta}
    &=
    \sum_i
    \Bigg[
        \frac1{4}
        \delta_{\alpha\beta}
        -
        c^4\overline{n}_{i,\alpha}  \overline{n}_{i,\beta} 
        -
        (1-c^2)^2 n_{i,\alpha}  n_{i,\beta} \nonumber\\
        &+
        \frac{\rm i}{2}\varepsilon_{\alpha\beta\gamma}(c^2\overline{n}_{i,\gamma} 
        +
        (1-c^2) n_{i,\gamma} )
    \Bigg],
\end{align}
where we omitted the terms that vanish regardless whether $c=0$ or $c=1$.
Moreover, due to the symmetric term $\delta_{\alpha\beta}-q_\alpha q_\beta$ in the cross section \eqref{eq:cross section}, the antisymmetric complex term does not contribute to the final observable.
These structure factors do not depend on the wave vector $\bm q$ and only contribute to the background.
Unlike the response of the skyrmion and antiskyrmion MPS, the connected cross sections of their product-state counterparts therefore show no significant structure.
The cross sections of the individual product states and their symmetric superposition are shown in \cref{fig:ps_sans}.
The conical structure of the cross sections is present merely because of the term $q_\alpha q_\beta$ in the definition \eqref{eq:cross section}.
%

%%%%%%%%%%
\section{Skyrmion dynamics in the classical limit}
\label{sec:thiele}
%%%%%%%%%%

\begin{figure}[ht!]
    \centering
    \includegraphics[width=\columnwidth]{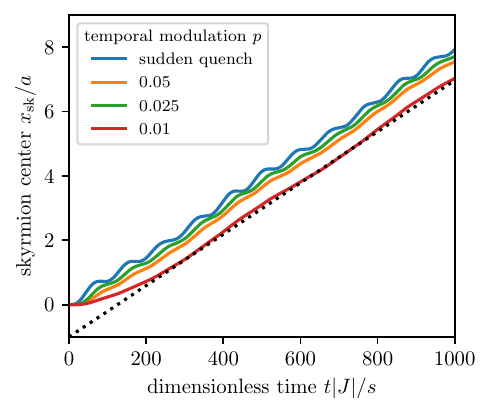}
    \caption{
        Trajectories of the skyrmion center relaxed in a monoaxial chiral magnet under a gradient field perturbation.
        As the modulating function $\frac{2}{\pi}\arctan(\sinh(pt|J|/s))$ increases more slowly, the velocity approaches the analytical solution \eqref{eq:velocity}.
    }
    \label{fig:x_com_t}
\end{figure}

%
Consider the following Hamiltonian with a perturbation in the form of a coordinate-dependent Zeeman term:
\begin{align}
    s\hat{H}
    &=
    -\frac{1}{2}\sum_{\langle ij\rangle}
    \left[
        J\hat{\bm{S}}_i\cdot\hat{\bm{S}}_j
        +
        \bm{D}_{ij}\cdot\left(\hat{\bm{S}}_i\times\hat{\bm{S}}_j\right)
    \right]
    \nonumber\\
    &\qquad -
    \sum_i
    \left(
        \bm{B}+\bm{B}_i^\mathrm{grad}
    \right) \cdot \hat{\bm{S}}_i
    ,
    \label{eq:Hamiltonian bare}
\end{align}
where $\bm{B}_i^\mathrm{grad}=B_\mathrm{grad}\frac{2}{N_y-1}(0,0,y_i)$ is a gradient field with a small amplitude $\vert B_\mathrm{grad}\vert \ll \vert B \vert $.
We employ the mean-field approximation to study how this perturbation induces the movement of topologically charged magnetic textures.
With the assumption that the spins of the system are mutually weakly correlated, we construct an effective Hamiltonian by introducing the spin fluctuation from the mean value, $\delta \hat{\bm{S}}_i = \hat{\bm{S}}_i - \bm{m}_i$, where $\bm{m}_i=\langle \hat{\bm{S}}_i\rangle$.
Plugging this into $\hat{H}$ and omitting second-order fluctuation terms, we obtain the mean-field noninteracting Hamiltonian:
\begin{align}
    s\hat{H}\approx s\hat{H}_\mathrm{MF} 
    &=
    -\sum_{\langle ij \rangle}
    \left[
        J\delta\hat{\bm{S}}_i \cdot \bm{m}_j 
        +
        \bm{D}_{ij} \cdot \delta\hat{\bm{S}}_i \times \bm{m}_j
    \right]
    \nonumber \\
    &\quad
    -\sum_i
    \left(
        \bm{B}+\bm{B}_i^\mathrm{grad}
    \right) \cdot \delta\hat{\bm{S}}_i
    +
    sE[\bm{m}],
\end{align}
where the term $sE[\bm{m}]$ is proportional to the energy of the mean magnetization texture $\bm{m}$.
As this term is not operator-valued, it does not contribute to the spin dynamics and may be omitted.
Furthermore, we assume that for negligible fluctuations, the texture has a uniform norm $\vert \bm{m}\vert = s$.
The time evolution of the spin operator $\hat{\bm{S}}_i$ in the Heisenberg picture is then determined via the commutator with the mean-field Hamiltonian,
\begin{align}
    \frac{\rm d}{\mathrm{d}t}\hat{\bm{S}}_i
    &=
    \frac{i}{\hbar}\left[\hat{H}_\mathrm{MF},\hat{\bm{S}}_i\right] = - \hat{\bm{S}}_i \times \bm{H}_i^\mathrm{eff}/s,
    \\
    \bm H_{i}^{\rm eff}
    &=
    - \sum_{j \in \mathrm{NN}(i)}
    \left[
        J\bm{m}_j + \bm{m}_j \times\bm{D}_{ij}
    \right]
    \nonumber\\
    &\qquad\qquad
    -\bm{B} - \bm{B}_i^\mathrm{grad},
\end{align}
For the local observables it follows that
\begin{equation}
    \dot{\bm{m}}_i
    =
    - \bm{m}_i \times \bm{H}_i^\mathrm{eff}/s,
\end{equation}
which is the Landau-Lifshitz equation.
Using the above equation, we can numerically resolve that topologically charged (anti)skyrmion domains display translational motion, with velocity perpendicular to the perturbation gradient.
The Thiele equation reads
\begin{equation}
    \bm{G}\times\bm{v} - \bm{F} =0, 
\end{equation}
where the gyrocoupling vector $\bf G$ and the force vector $\bf F$ are defined as
\begin{align}
    G_{i}
    &=
    - \varepsilon_{ijk} \frac{1}{2s^2}
    \int\left[
        \bm{m}
        \cdot
        \left(
            \frac{\partial\bm{m}}{\partial x_j}
            \times
            \frac{\partial\bm{m}}{\partial x_k}
        \right)
    \right]\,dxdy
    , \\
    F_i
    &=
    \frac1s
    \int\left[
        \bm{H}^\mathrm{eff}
        \cdot
        \frac{\partial\bm{m}}{\partial x_i}
    \right]\,dxdy.
\end{align}
For a skyrmion that is considered to be rigid, the internal forces related to exchange, Dzyaloshinskii-Moriya, and Zeeman interactions must be zero.
The gradient field contributions for a $L_x \times L_y$ rectangular flake are
\begin{align}
    &F_x
    =
    \frac{2B_\mathrm{grad}}{sL_y}
    \int
    y \frac{\partial m_z}{\partial x}
    d^2r 
    =
    \frac{2B_\mathrm{grad}}{sL_y} \times \\
    &\times 
     \int\limits_{-L_y/2}^{L_y/2} y\left[m_z\left(\frac{L_x}{2},y\right)-m_z\left(-\frac{L_x}{2},y\right)\right]dy 
    = 
    0,
    \nonumber
\end{align}
and
\begin{align}
    &F_y = \frac{2B_\mathrm{grad}}{sL_y}\int y \frac{\partial m_z}{\partial y} d^2r
    =
    \frac{2B_\mathrm{grad}}{sL_y} \times \nonumber\\
    &\times
    \int   \underbrace{\left[\frac{L_y}{2}m_z\left(x,\frac{L_y}{2}\right)+\frac{L_y}{2}m_z\left(x,-\frac{L_y}{2}\right)\right]}_{sL_y} dx \nonumber \\
    &\qquad\qquad- 
    \frac{2B_\mathrm{grad}}{sL_y}\int m_z d^2r.
\end{align}
We considered the edge magnetization to be fixed along the direction of the Zeeman field, $\bm{m}\vert_\mathrm{edge}=s\bm{B}/\vert \bm{B}\vert=(0,0,s)$.
Equivalently, semiperiodic or periodic boundaries may be used.
The non-vanishing force becomes
\begin{align}
    F_y&= - \frac{2B_\mathrm{grad}}{sL_y} \int (m_z-s) d^2r  \equiv - \frac{2B_\mathrm{grad}}{sL_y} N,
\end{align}
where $N$ is the total deviation of $m_z$ from the amplitude $s$.
For a domain with a rigid shape, $N$ is constant in time. 
As the $z$-derivatives of the magnetization vanish, the only nonzero element of the gyro-vector is $G_z$,
\begin{equation}
 G_z= \frac{1}{s^2}\int  \left[ \bm{m} \cdot\left(\frac{\partial\bm{m}}{\partial y}\times\frac{\partial\bm{m}}{\partial x}\right)\right]dxdy = -4\pi Q s,
\end{equation}
where we used the formula for the topological charge defined for the unit vector field $\bm{n}=\bm{m}/s$,
\begin{equation*}
    Q = \frac{1}{4\pi}\int  \left[ \bm{n} \cdot\left(\frac{\partial\bm{n}}{\partial x}\times\frac{\partial\bm{n}}{\partial y}\right)\right]dxdy.
\end{equation*}
The Thiele equation then becomes,
\begin{equation}
    \begin{pmatrix}
        4\pi Qsv_y \\ -4\pi Qsv_x \\ 0
    \end{pmatrix} =
    \begin{pmatrix}
        0 \\ -2B_\mathrm{grad}N/(sL_y) \\ 0
    \end{pmatrix},
\end{equation}
with the solution 
\begin{equation}
    \bm{v}= \frac{B_\mathrm{grad}N}{2\pi s^2 Q L_y }\bm{e}_\mathrm{x}. \label{eq:velocity}
\end{equation}
The simulated velocity of the skyrmion can be compared with the analytical result \eqref{eq:velocity} obtained for the continuum limit.
We observe that the skyrmion trajectory displays oscillations about an average linear path in both the classical and quantum simulations.
This comes from the sudden quench of the gradient field, as it is effectively introduced as $B_\mathrm{grad}(t)\propto \Theta(t)$.
We conclude that an abrupt jump in the amplitude of the perturbation excites other skyrmionic motion modes such as breathing or rotation modes on top of the linear translation.
In \cref{fig:x_com_t}, we show that the oscillations level out for $B_\mathrm{grad}(t)$ modulated by a slowly increasing smooth function.
We choose to parametrize the gradient field with $\left(2/\pi\right)\arctan(\sinh(pt|J|/s))$ and vary $p$.
\\  % this linebreak is necessary to avoid cropping of text due to a glitch in placing the next floating figure

%

%%%%%%%%%%
\section{Convergence of TDVP simulations}
\label{sec:tdvp}
%%%%%%%%%%

\begin{figure*}[ht]
    \centering
    \includegraphics[width=\linewidth]{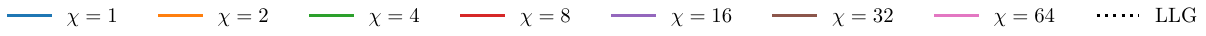}
    \\
    \includegraphics{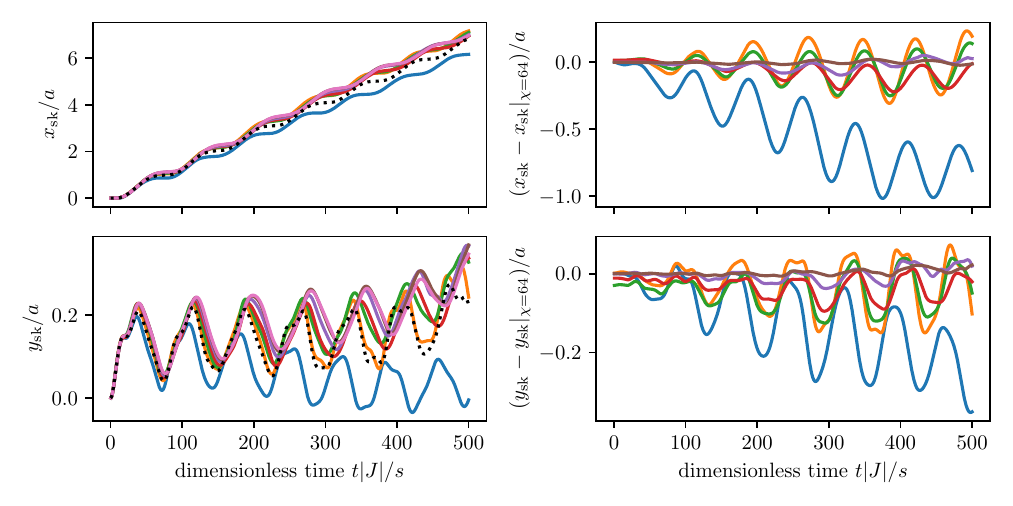}
    \caption{
        TDVP results of the skyrmion motion $x_{\rm sk}$ and $y_{\rm sk}$ for different bond dimensions $\chi$.
        For $\chi=32$, the lines visually collapse with $\chi=64$ over a large time window, with a maximum deviation $0.034a$ in $x_{\rm sk}$ and $0.022a$ in $y_{\rm sk}$, respectively.
        The dashed line corresponds to the result from classical LLG simulations.
    }
    \label{fig:tdvp_different_M}
\end{figure*}

%
We consider the dynamics of the approximate eigenstates of $\hat H$ by simulating the unitary evolution generated from the perturbed Hamiltonian $\hat H$.
The best-effort converged DMRG-X states with $\chi=16$ are used as the initial states of the TDVP algorithm.
During simulation, we allow the bond dimension to increase to a maximum of $\chi=64$.
We exploit the exact relation $\ket{\psi_{\rm ask}(t)} = \hat K\ket{\psi_{\rm sk}(-t)}$ to simulate the antiskyrmion trajectory via the time-reversed skyrmion trajectory.
The results for different bond dimensions are presented in \cref{fig:tdvp_different_M}.
We find a visible collapse of the $x_{\rm sk}$ trajectory for $\chi\in\{16,32,64\}$ in the total time window, with a difference that consistently decreases when $\chi$ is increased.
This trend is similar for $y_{\rm sk}$.

\end{document}